\documentclass[12pt]{article}
\input epsf.tex

\usepackage{latexsym}
\usepackage{graphicx}
\usepackage{amsmath,amstext}
\usepackage{comment}
\numberwithin{equation}{section}
\newcommand{\be}{\begin{equation}}
\newcommand{\ee}{\end{equation}}
\newcommand{\benn}{\begin{equation*}}
\newcommand{\eenn}{\end{equation*}}
\newcommand{\bea}{\begin{eqnarray}}
\newcommand{\eea}{\end{eqnarray}}
\newcommand{\bean}{\begin{eqnarray*}}
\newcommand{\eean}{\end{eqnarray*}}

\begin{document}
\begin{titlepage}
\begin{center}

\hfill SCIPP 08/13\\

\vskip0.5in

{\Large \bf Surveying the Phenomenology of General Gauge Mediation}

\vskip 0.3 in

Linda M. Carpenter$^1$

\vskip0.2in

\emph{$^1$ University of California Santa Cruz\\Santa Cruz CA\\ lmc@scipp.ucsc.edu}

\begin{abstract}
I explore the phenomenology, constraints and tuning for several weakly coupled implementations of multi-parameter gauge mediation and compare to
minimal  gauge mediation.  The low energy spectra are distinct from that of minimal gauge mediation; a wide range of NLSPs is found and spectra are
significantly compressed, thus tunings may be generically reduced to a part in 10 to a part in 20.

\end{abstract}
\end{center}
\end{titlepage}

\section{Introduction}
Gauge mediation is a predictive and flavor blind communication mechanism for SUSY breaking \cite{Dine:1993yw}.
The earliest and simplest implementation of minimal gauge mediation involved communication of SUSY breaking from the hidden sector through a single set
of vector like messengers charged under standard model gauge groups.
The messengers couple to a hidden sector field $\langle X \rangle = x + \theta^2 F_x$ and acquire a supersymmetric and nonsupersymmetric mass term,

\be
W = \lambda X  \phi \tilde \phi  \rightarrow  \lambda \langle x \rangle \phi \tilde \phi + \lambda F_x \phi \tilde \phi
\ee

Defining $\Lambda \equiv F_x/x$ we see that one loop gaugino masses are generated

\be
M_{\lambda i} \sim \frac{\alpha_i}{4\pi}\Lambda
\ee

in addition to two loop scalar masses
\be
\widetilde m^2 ={2 \Lambda^2} \left[C_3\left({\alpha_3 \over 4 \pi}\right)^2
+C_2\left({\alpha_2\over 4 \pi}\right)^2
+{5 \over 3}{\left(Y\over2\right)^2}
\left({\alpha_1\over 4 \pi}\right)^2\right].
\label{scalarsmgm}
\ee

Minimal gauge mediation predicts that all scalar and gaugino mass terms are related by a single mass scale.  Sparticles get masses proportional to
powers of their gauge couplings, therefore, there is a large mass hierarchy in the spectrum placing sparticles charged under QCD far above the others.
 A large mass hierarchy in the SUSY spectrum is not just a prediction of gauge mediation but also results from the other leading candidates for SUSY
 communication, anomaly mediation and mSUGRA \cite{Randall:1998uk}.  Recent bounds on quickly decaying NLSPs place a strong limit on $\Lambda$, the mass scale
 of minimal gauge mediation, at 91 TeV \cite{:2007is}.  For non-prompt NLSP decay lower bounds come from charged sparticle masses and
 indirect constraints such as the inclusive tri-lepton signal.  In both cases a heavy and thus tuned spectrum results,
 but this applies especially to the case of prompt NLSP decay.  As experiments push the spectrum of minimal gauge mediation higher, non-minimal models
  become more phenomenologically (and for many philosophically) attractive.

Simple extensions to MGM have been scattered throughout the literature, see for example \cite{Dimopoulos:1996yq} \cite{Kribs:1999df}.
Recently \cite{Meade:2008wd} have proposed a simple framework for counting allowed parameters in general gauge mediated schemes.   The
general framework allows for at most six mass parameters to determine the low energy spectrum.  There are now a variety of viable models
implementing GGM in a more systematic way, however a detailed analysis of the parameter space of non-minimal SUSY models has not been done.  Though
 constraints on minimal frameworks like MSUGRA are well understood, in non-minimal scenarios it isn't even clear exactly what bounds on sparticle
 masses are - for example see recent work on gluinos \cite{Alwall:2008va} or look through PDG to see how many analyses are solely MSUGRA based\cite{pdg}.


For a given low energy spectrum in the GGM framework, there may be several ways to complete the model in the hidden sector.  Model building in general
gauge mediation may be accomplished with weakly coupled renormalizable theories \cite{Cheung:2007es} \cite{Carpenter:2008wi},
or in frameworks with non-perturbative dynamics \cite{Seiberg:2008qj} \cite{Carpenter:2008rj}.  GGM models may be direct, with messengers
participating in SUSY breaking, or indirect.  In fact the general framework of GM does not require messengers at all.   The goal of this work is not
to build specific UV completions to models, or to insist on a specific fix in the Higgs sector.  Rather it is to look at
the low energy predictions and pick out interesting regions of parameter space where the phenomenology is far from standard and/or the tuning is significantly less then the minimal case.

That said, many implementations of GGM result in split SUSY like spectra, where scalars are much heavier than gauginos and tuning is quite bad.
This may be the case for example if Poppitz-Trivedi type mass terms are large.  Such terms add to scalar masses only as they preserve
R symmetry and are accounted for by non-vanishing messenger supertrace \cite{Poppitz:1996xw}.  Such terms occur generically in certain classes of GGM models.  This work will not
address models with large mass splitting between scalars and gauginos, rather it will attempt to focus on theories that predict light spectra which look
different from both the MGM and split SUSY spectra.  The focus here will be limited the predictions of weakly coupled renormalizible hidden sectors
with and R symmetry \cite{Carpenter:2008wi}.  Such models avoid split spectra and rely on multiple hidden sector spurions and different numbers of messenger multiplets in various representations of SU(5) to change the number of low energy parameters.

 For example, consider the case of N sets of messengers in a $5$, $\overline{5}$ representation of SU(5).  These messengers may couple to multiple scalar fields which get vevs and F terms.  The messenger couplings are

\be
\lambda_\ell^{a i} X_a \bar \ell_i\ell_i + \lambda_q^{a i} X_a \bar q_i q_i
\ee
where the index i counts the N messengers and the index a counts the hidden sector fields.

There are then four parameters which determine the low energy spectrum,

\be
\Lambda_q^i = {\lambda_q^{a i} F_a \over \lambda_q^{b i} x_b}; \Lambda_\ell^i = {\lambda_\ell^{a i} F_a \over \lambda_\ell^{b i} x_b}.
\ee

Counting this way we see that one set of $5$,  $\overline{5}$ leads to two parameters, one set of $10$, $\overline{10}$ yields three and so on.  Scanning
over the complete space of gauge mediated parameters is complicated.  However, the possible low energy phenomenologies become very nonstandard
for cases with just two or three parameters.  This work will focus on such cases, with the hope that these scans will sample much of the interesting low energy phenomenology of the full six parameter space.  What follow in section 2 is an overview of parameter
counting for weakly coupled GGM models.  Section 3 is a catalogue of the most relevant constraints on GGM parameter space.  The remaining sections map out
the viable parameter space for different multi-parameter GM scenarios; section 4 deals with 2 parameter GM, section 5 with 3 parameters.
Section 6 concludes.

\section{Counting Parameters and Using Benchmarks}

In counting parameters we must of course first count the number of scales $\Lambda_i$ which determine the gauge mediated soft mass spectrum.  There is
however a difference between the number of parameters predicted by general gauge mediation and the number of parameters involved in producing a viable
 spectrum.

Models within the gauge mediated framework must include additional structure that generates  $\mu$ and $B_{\mu}$ terms since these parameters do not
fall within the definition of gauge mediation, $\it{i.e.}$ these terms occur wether the SM gauge couplings are turned off or not.  However, any possible
completion that generates $\mu$ may also effect Higgs soft masses.  Therefore even though GGM makes Higgs soft mass predictions at each point in
 parameter space, the soft masses must be thought of as parameters to be scanned over.  Contributions to soft masses squared
 may either be positive or negative, therefore we must consider benchmark scenarios, the most conservative assuming as small a
 positive contribution to Higgs soft masses as possible, as well as scenarios assuming large negative mass squared contributions.
  The parameters in this sector are the b-term, $\mu$, tan $\beta$,and the soft masses $m_{Hu}$ and $m_{Hd}$.  For the purposes of this paper
   all scans are done fixing tan$\beta$ at 10, once the soft masses are selected, $\mu$ and $B_{\mu}$ are solved for using the conditions of the
    electroweak minimum.

The SUSY breaking scale also effects the spectrum and experimental constraints.  The SUSY breaking scale controls the mass of the gravitino,
the LSP in gauge mediation, and determines the lifetime of sparticles.  As the SUSY breaking scale is varied, sparticles may decay
inside or outside of the detector drastically changing the phenomenological constraints.  In addition, the sparticle masses run from the
SUSY breaking scale altering the low energy spectrum.  Two benchmark scenarios will be considered; for the case of prompt sparticle decay the
low SUSY breaking scale is taken to be $10^5$ GeV,  for decay outside of the detector the high SUSY breaking benchmark is chosen to be $10^{8}$ GeV.

The number of copies of messengers effects the spectrum but requires some caution.  In MGM the gaugino masses
scale like N the number of messengers, and the scalar mass squareds scale like the $\sqrt{N}$.  The ratio of scalar to gaugino masses may thus be
compressed by a factor of $\sqrt{N}$.  This has significant consequences for searches, for example, large N may make the difference between spectra
with a bino-like LSP and a stau LSP.  However increasing N may itself alter the number of GGM parameters.
  An example will illustrate the point.  In the case of two parameter GM, achieved with one messenger set in a 5 of SU(5), the masses are given by the 2D subset of equation 1,
\be
m_{\tilde{g}}=\frac{\alpha_3}{4\pi}\Lambda_3;\quad m_{\tilde{w}}=\frac{\alpha_2}{4\pi}\Lambda_2;\quad m_{\tilde{b}}=\frac{\alpha_1}{4\pi}(\frac{2}{3}\Lambda_3+\Lambda_2)~;
\ee

\be
m_{\tilde{s}}^2=2\left(C_3 (\frac{{\alpha_3}}{4\pi})^2 \Lambda_3^2 +C_2 (\frac{\alpha_2}{4\pi})^2 \Lambda_2^2 +\frac{Y}{2}^2(\frac{\alpha_1}{4\pi})^2 (\frac{2}{3}\Lambda_3^2 + \Lambda_2^2)\right)~.
\ee
What happens if we want two copies of the messengers?  Generally this will not just multiply these equations by N, but will instead put us in a four dimensional parameter space,

\be
m_{\tilde{g}}=\frac{\alpha_3}{4\pi}(\Lambda_3+\Lambda_3^{'});\quad m_{\tilde{w}}=\frac{\alpha_2}{4\pi}(\Lambda_2+\Lambda_2^{'});
\ee
\be
m_{\tilde{b}}=\frac{\alpha_1}{4\pi}(\frac{2}{3}(\Lambda_3+\Lambda_3^{'})+(\Lambda_2+\Lambda_2^{'})) \nonumber;
\ee

\bea
m_{\tilde{s}}^2&=&2\left(C_3 (\frac{\alpha_3}{4\pi})^2 (\Lambda_3^2+\Lambda_3^{'2}) +C_2 (\frac{\alpha_2}{4\pi})^2 (\Lambda_2^2 +\Lambda_2^{'2})\right. \nonumber \\
& &\left.+\frac{Y}{2}^2(\frac{\alpha_1}{4\pi})^2 (\frac{2}{3}(\Lambda_3^2+\Lambda_3^{'2})+(\Lambda_2^2+\Lambda_2^{'2})\right)~.
\eea
Where the four parameters are the two sums of scales $\Sigma \Lambda_i$ and sums of the square of scales $\Sigma {\Lambda_i}^2$.  This counting will be true for any number or messenger multiplets beyond two.  We see that in our example there is a 2-D subspace of this 4 dimensional space
where $\Lambda_i=\Lambda_i^{'}$ which would multiply both the scalar mass squareds and the gaugino masses by 2.

Finally the predicted Higgs mass must exceed the experimental bound.  For standard signals LEP puts a lower mass
 bound on the Higgs of 114 GeV \cite{higgs bound}.  However the MSSM predicts a tree-level Higgs mass of at most the mass of the Z.
 There are several ways to solve this problem.  The standard solution is to include loop corrections from very heavy
 stops.  At one loop the stop correction is overcounted, at
two loops a stop of 500 GeV gives a Higgs mass of roughly 105 GeV \cite{Haber:1996fp} \cite{Carena:2000dp}.  In order to push the Higgs above
the mass bound much heavier stops are needed.  However, loop corrections from heavy stops also make large contributions to $m_{Hu}$, which must the be
canceled down to $m_Z$ in the electroweak potential. For this reason all models become tuned to roughly one part in $\delta m_{Hu}^2 / m_Z^2$.  Barring the heavy stop solution,
models may be completed with a variety of other fixes. The first is by assuming a hidden Higgs scenario, where the Higgs is in fact lighter
than 114 GeV, but decays in a nonstandard way so as to have escaped detection.  This may happen in R parity violating scenarios \cite{Carpenter:2008sy} or in the NMSSM\cite{Dermisek:2005ar}
.
The second is by assuming extra operators in the Higgs sector, for example the dimension 5 operators proposed by Dine Seiberg and
Thomas \cite{Dine:2007xi}.  These operators alter the calculation of the electroweak minimum and present one or two extra parameters.
In particular dimension 5 operators change the Higgs potential in the following way

\be
\delta_1 V + \delta_2 V = 2 \epsilon_1 ~
   H_u H_d ~( H_u^\dagger H_u + H_d^\dagger H_d ) + \epsilon_2 ~(H_u H_d)^2   ~+~{\rm h.c.}
 \label{vepsilon1} \ee
 where
  \be \epsilon_1 \equiv {\mu^* \lambda \over M},  \epsilon_2 = - {m_{\rm SUSY} \lambda \over M} . \ee

and thus add to the Higgs mass
\be
\delta_\epsilon m_h^2 = 2v^2\left(\epsilon_{2r} + 2 \epsilon_{1r}
 \sin(2\beta) + {2 \epsilon_{1r}(m_A^2 + m_Z^2) \sin(2\beta) -
 \epsilon_{2r} (m_A^2-m_Z^2) \cos^2(2\beta) \over
 \sqrt{(m_A^2-m_Z^2)^2 + 4m_A^2 m_Z^2 \sin^2(2\beta)}} \right)
 \ee

This does not alter the condition for electroweak minimum by much but it does call for an intermediate scale of physics at a TeV.  As per
the calculation of Dine {\it et. al.},  at $tan\beta$ 10 and choices of $\lambda/M$ of order $10^{-4}$, The Higgs mass may be set above
LEPs bound for squark masses of 300 GeV.  In this work, all scans are done assuming a value $\lambda/M = 2$x$10^{-4}$.

\section{Constraints}

\subsection{Direct Mass Bounds}

\begin{table}[h]
\begin{center}
\begin{tabular}{c|c|c}
\hline
gauginos&  prompt decay & non-prompt decay \\
\hline
bino-like NLSP & & \\
\hline
$\chi^{+}$  & 229 & 102.7 \\
$\tilde g $  & 320 & 130 \\
\hline
gluino NLSP & &  \\
\hline
$\tilde g $  & 315 & 270 \\
\hline
\end{tabular}
\end{center}
\end{table}
\begin{table}[h]
\begin{center}
\begin{tabular}{c|c|c}
\hline
wino-like NLSP &  $m_{\chi^{+}}- m_{\chi^{0}} < m_{\pi}$ & $ m_{\pi} < m_{\chi^{+}}- m_{\chi^{0}} < $ 3GeV \\
\hline
$\chi^{+}$  & 206 & $45^{\dagger}$ \\
\hline
\end{tabular}
\caption{direct mass bounds on gauginos decaying through different NLSPs, $\dagger$ assumes light sneutrino masses }
\label{tab:gauginobounds}
\end{center}
\end{table}

Tevatron places strong inclusive bounds on charged massive stable particles, or CHAMPS, which must be taken into account in configurations where the LSP
is both long lived and charged.  Bounds on CHAMPS are quite restrictive, the upper limit on production for weakly interacting particles  is 10 $fb$ \cite{Abazov:2008qu}.

Direct bounds on gauginos depend strongly on the NLSP, and if the SUSY breaking scale is low or high.  A recent redux of bounds is found in
reference \cite{Berger:2008cq}.  The absolute lower limit on chargino
masses, which is independent of decay products, comes from LEP 1 and is 45 GeV.  In the case of a bino-like NLSP  and sufficient mass splitting LEP
places a chargino mass bound of
102.7 GeV \cite{Abdallah:2003xe}.  In the case that the NLSP is wino-like, the mass splitting between the chargino and lightest neutralino may be
 very small.  If it is under 3 GeV, the previously stated bound no longer holds.  If the mass splitting is less than the pion mass then the chargino lives a
 very long time \cite{Gunion:1999jr} and its existence may be ruled out by CHAMP data. The lower limit on long lived stable charginos
 is 206 GeV \cite{Abazov:2008qu}.  However LEP's analyses also assume a small contribution from t channel sneutrinos, if sneutrinos are light the bounds do not apply .
  In the case of low SUSY breaking scale such that the chargino decays inside the detector, the mass bound is a very
restrictive 229 GeV \cite{:2007is}.

A decay mode independent lower mass bound on the gluino has been set at 51 GeV \cite{Kaplan:2008pt}.  In regions of  parameter space that obey MSUGRA
like relations, the gluino can decay through real or virtual squarks to a bino-like NLSP.  Recent work has placed a strict lower bound on the
 gluino decaying to jets plus missing energy with nonunified gaugino masses at 130 GeV \cite{Alwall:2008va}, and has also made
 exclusions in the gluino-bino mass plane.  Previous searches for gluinos had been confined to the MSUGRA scenario, where the current PDG
 lower bound is quoted at 308 GeV \cite{pdg}.  MSUGRA assumes that the gluino decays to a bino LSP which is much lighter than the gluino
 itself, which would mean a substantial amount of missing energy in the event.  However in non MSUGRA scenarios the amount of missing energy decreases
 as the gluino mass and bino mass are moved closer together.  Therefore the Tevatron's missing energy events cuts would cause gluinos to be missed for
 a range of masses.  Recent work not only moves the MSUGRA bound up from 308 GeV, but it makes exclusion in the full gluino-bino mass plane.
 These bino vs. gluino mass constraints will be considered in the following scans.  In the case of prompt decay,
 if the bino is the NLSP the experimental constraints are stronger; there is a search for gluinos decaying to squarks
 plus diphotons and missing energy which placed the gluino mass bound at 320 GeV \cite{Abbott:1998gi}.  This search had relatively low cuts on
 missing energy and thus it is not possible to avoid constraints.
  In the case that the gluino is the NLSP, non-prompt decay of the stable gluino is ruled out by stopped gluino searches to at least 270 GeV for
 particle lifetimes under 3 hours\cite{Abazov:2007ht}.  For prompt decay Tevatron's monojet search rules it out to 320 GeV \cite{Abulencia:2006kk}.

\subsection{Indirect and Other Constraints}
The trilepton signal is a distinctive SUSY signature.  Tevatron has done searches for the final state: three leptons plus missing
energy \cite{Glatzer:2008ur}.  In the SUSY framework this is a signal for the decays of a charged and neutral gaugino.
The process is $\chi^{+} \rightarrow \tilde{l} +\nu, \chi^{0}_2 \rightarrow \tilde{\l}+ l$ where $\tilde{l}$ decays to lepton  and $\chi^{0}$.
Bounds on the trilepton cross section are extremely small.  There is room for half an event at two inverse $fb$ of data.  Thus if
 $\epsilon \sigma (bf) L = N$ we see that even for a low efficiency, for example 5 percent, there is only room for $fb$'s of
 production.  There are three possible outs: first, we may make the gauginos heavy to reduce the production cross section.  Second we may significantly
 decrease the branching fraction into leptons and missing energy.  For example if the intermediate sleptons are far off shell, the decay will instead
 proceed through virtual gauge bosons; in this case the branching fraction to leptons plus missing energy is lowered because of significant contribution
 from branching fraction to jets. Finally, the chargino may be made Higgsino like, in which case the final decay will favor taus, excluded by the
 trilepton analysis.  The triletpton search highly constrains any model which predicts dominant decays of charginos and neutralinos which proceed
 through sleptons, e.g. in regions of SUSY parameter space where the neutralino is the NLSP and the sleptons are light, which is exactly the case in
 minimal gauge mediation with non-prompt NLSP decay.   This bound puts a new lower limit on the mass of the lightest chargino beyond the direct searches.  The new bound for charginos in the MSUGRA scenario is 145 GeV \cite{Dube:2008kf}
and is expected to be 160 GeV with 12 $fb^{-1}$ of data.  One expects that with one fewer parameter, minimal gauge mediation will have a bound at least
about as strict.  For GGM the bound will depend on the spectra and will be a strong constraint for models with certain LSPs and mass hierarchies.

The Higgs sector conditions of electroweak minima are well known: any model must reside in a region of SUSY parameter space where the electroweak
symmetry is broken and the Higgs acquires the proper vev.  The parameters in this sector are the b-term, $\mu$, tan $\beta$,and the soft
masses $m_{Hu}$ and $m_{Hd}$ but they are not all independent, the conditions of the electroweak minima remove two of these parameters
(also recall there are two additional inequalities which are constraints for the stability of the potential).

The mass of the charged Higgs, the charginos and the stops are predicted at each point in parameter space.   These parameters are
involved in the prediction of the process $b \rightarrow s\gamma$.  There is a tight bound on this process \cite{Barberio:2007cr}.  However, the
NNLO prediction is for the first time below the measured value, leaving some space for SUSY corrections \cite{Misiak:2006zs}.  Scans are conducted over
all parameter space calculating  $\Gamma (b\rightarrow s\gamma)$ at each point.

Finally the Higgs mass must be sufficiently increased.

\section{Two parameter GM}
\begin{figure}[h]
\centerline{\includegraphics[width=10 cm]{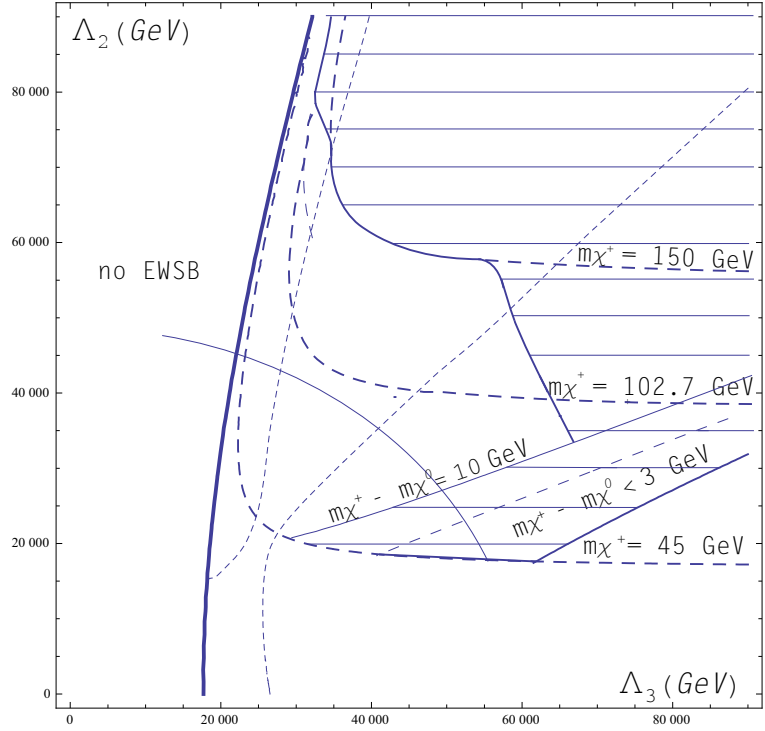}}
\caption{Plot of $\Lambda_3$ vs $\Lambda_2$ with various constraints.  Allowed parameter space is shaded. The arc on the bottom left is LEP's right handed selectron bound, the dotted line indicates where the mass difference between the chargino and the selectron is zero.}
\label{fig:2pgm}
\end{figure}

Consider a two dimensional GGM parameter space achieved with a set of messengers in the 5, $\overline{5}$ representation.  The two GM parameters are
the scales $\Lambda_2$ and $\Lambda_3$ and the sparticle spectrum is given by
\be
m_{\tilde{g}}=\frac{\alpha_3}{4\pi}\Lambda_3;m_{\tilde{w}}=\frac{\alpha_2}{4\pi}\Lambda_2;m_{\tilde{b}}=\frac{\alpha_1}{4\pi}(\frac{2}{3}\Lambda_3+\Lambda_2)
\ee

\be
m_{\tilde{s}}^2=2(C_3 (\frac{\alpha_3}{4\pi})^2 \Lambda_3^2 +C_2 (\frac{\alpha_2}{4\pi})^2 \Lambda_2^2 +\frac{Y}{2}^2(\frac{\alpha_1}{4\pi})^2 (\frac{2}{3}\Lambda_3^2 + \Lambda_2^2)
\ee

In the case that $\Lambda_3 = \Lambda_2$ we are reduced to the minimal gauge mediated relations.  We see immediately that lower $\Lambda_3$ in
relation to $\Lambda_2$ may compress the SUSY spectrum, while raising the ratio will make the QCD hierarchy even bigger than the minimal scenario.
In parts of the parameter space where $\Lambda_3$ is sufficiently less than $\Lambda_2$ the gluino can be made lighter than the bino, making it the NLSP.
In regions where $\Lambda_3$ is sufficiently larger than $\Lambda_2$, then $m_{\tilde{w}}$ $< m_{\tilde{b}}$, and the NLSP is wino-like.  Everywhere
else in parameter space, the NLSP is bino-like.

The MGM prediction of heavy squarks relative to the sleptons does not necessarily hold.  Because the sleptons hypercharge is larger than
that of up type squarks
by a factor of 3, one finds that for regions of mass parameters such that $\frac{3}{2}\frac{\alpha_3}{\alpha_1}^2\Lambda_3^2<\Lambda_2^2$ the sleptons are
heavier than the squarks.

We now concentrate on the constraints on parameter space in the high SUSY breaking scale scenario.
The first constraint is to satisfy the conditions for the electroweak minimum.  In the case that $m_{Hu}$ does not get
large negative mass contributions from the $\mu$ generating mechanism, this requires $m_{Hu}$ to run down due to the stop coupling. This in turn
sets a lower
limit on squark masses and hence $\Lambda_3$. Extra contributions to $m_{Hu}$ will shift the EWSB constraint around the parameter space, increasing or
decreasing the minimum allowable $\Lambda_3$ and altering the size
of the $\mu$ term as $\mu$ cancels the
soft Higgs mass in the minimization conditions of the electroweak potential.  For the choice of dimension five Higgs sector operator, the Higgs mass remains above 110 and
114 GeV for stop masses of 310 and 360 GeV respectively.  Figure 1 gives a picture of parameter space for the benchmark scenario
where the Higgs soft masses are set at their gauge mediated soft mass predictions. We see that regions of low $\Lambda_3$ are ruled out,
including those regions where the gluino is the NLSP.

Of the remaining constraints, the most restrictive in the region of bino-like NLSP is the trilepton search.  Where the selectron is lighter
than the chargino and next to lightest neutralino, the branching fraction of a gaugino pair into three leptons plus missing energy is large.
Interpolating from the MSUGRA analysis we see that with 12 $fb$'s of data the lower limit should be around 150 GeV where the sleptons are light.  As
 the selectron mass is increased the branching fraction drops.  However if the masses of the chargino and lightest neutralino get too closer together,
 the resulting sleptons will be too soft.  The Tevatron's search required 4 GeV of $p_T$ for leptons \cite{Aaltonen:2008pv}.  Tracking the relaxation of the trilepton constraint due to small mass difference requires more rigorous analysis.
 In Figure 1 the strip of parameter space with mass difference between the chargino and lightest neutralino below 10 GeV is labeled.  This space
 may be ruled out by further analysis.

In addition, there is a lower bound from the model independent gluino search \cite{Alwall:2008va}.  However, the ratio of gluino to squarks mass is moderate
 and the bino mass constraint becomes less strict as
$\tilde{q} \tilde{g}$ production begins to compete with gluino pair production, therefore the model independent gluino mass bound only rules out a
small slice of parameter space.   In the region where $\Lambda_2$ is large, the chargino and bino masses are sufficiently split and the
minimum allowable chargino mass is 102.7 GeV.  However, in the region of
low $\Lambda_2$ the sneutrinos are quite light and for this reason a high chargino mass constraint does not apply.  Additionally
in regions of low $\Lambda_2$ compared to $\Lambda_3$, the wino-like neutralino is the NLSP.  Parameter space is allowed where the mass splitting
of the $\chi^{+}$ and $\chi^{0}$ is above the pion mass and where the $\chi^{+}$ is above 45 GeV.  For mass splittings smaller than the pion mass the chargino is long lived and ruled out by CHAMP
searches.

The regions of parameter space where the spectrum is most compressed are those with bino-like NLSP where $\Lambda_3 < \Lambda_2$.  A typical
spectrum follows,

\begin{table}[h]
\begin{center}
\begin{tabular}{c|c}
sparticle & mass (GeV)  \\
\hline
$\tilde{\tau}$  & 138  \\
$\tilde{\nu}$  & 246 \\
$\tilde{q_{ul}}$  & 494  \\
$\mu$  & 315  \\
$\chi^{+}$  & 158  \\
$\chi^{0}_1$  & 89  \\
$\tilde {g}$ & 343 \\
$ mz^2 / \delta_{mhu}^2$ & 20 \\

\end{tabular}
\caption{2 parameter spectrum, for $\Lambda_3$ 40 TeV, $\Lambda_2$ 65 TeV with Higgs soft masses set at the GGM prediction. $ mz^2 / \delta_{mhu}^2$ is the tuning measure.}
\label{tab:gauginobounds}
\end{center}
\end{table}

As a comparison, the minimal case satisfying all bounds requires squarks of around 700 GeV and a tuning of at least a part in 45. Then for this
simplest extension  of MGM
with a conservative benchmark scenario, the spectrum can be compressed by a factor of 1.5 and the tuning reduced by the square of that, though the
qualitative nature of the spectrum is similar to that of minimal gauge mediation.

In principle, if models exist with large negative contributions to the Higgs soft mass squared, the minimum allowable $\Lambda_3$ can be decreased.
For example, with models that add an additional $200$ $GeV^2$ to the up type Higgs mass, allowed up type squark masses can be
pushed below 250 GeV.  Here the Higgs mass is above 110 and 114 GeV for stop masses above 260 and 312 GeV.  The Tevatron's current squark search
limits the squark masses above 329 GeV, however PDG's bound for cascade decays where
the gluino mass is less than the squark mass is less restrictive, only 224 GeV \cite{pdg}.  One would expect a model independent
analysis to be slightly less restrictive than this as the missing energy of the event in the general case is less than the MSUGRA case since
the neutralino and gluino mass ratio
is compressed.  Note that in scenarios where the $m_{Hu}$ get a very large negative mass squared contribution for any Higgs mass fix, parameter
space is opened in the region gluinos NLSPs.  However a stable gluino NLSP is ruled out in the given region by bounds on stopped gluinos,
since gluino masses in this region are less than 200 GeV \cite{Abazov:2007ht}.  Beyond that the a lower limit on the squarks may be set due to the Higgs mass bound.  In general the Higgs mass of
110 GeV may only be surpassed for squark masses not much below 300 GeV, resulting in an irreducible tuning of 10 percent.

In the case of a promptly decaying NLSP, the parameter space is more severely restricted.
The charginos of mass bound of 229 GeV holds \cite{:2007is} and rules out much of parameter space, unless the mass splitting between chargino and
neutralino is small.  In this case having two parameters in gauge mediation is not sufficient to avoid tunings and restrictions on parameter space.




\subsection{The 2 Parameter Subspace of 4 Parameter GM}

If we take the case of messengers in a 5, $\overline{5}$ representation and increase the number of messenger multiplets, we are analyzing a 2-D subspace
of 4 parameter gauge mediation. Five 5, $\overline{5}$ messengers is more or less the maximum compatible with unification, in this case I have chosen to
map out the scenario with 4 messengers.  Here we get a factor of $\sqrt{4}$  between gauginos and scalars relative to the case with one
messenger multiplet.  Similar to MGM with multiple messengers, almost everywhere in parameter space the stau is the NLSP. As before the gluino
still has some small space to be the NLSP, and for
large values of $\Lambda_3$ compared to $\Lambda_2$ the NLSP is the wino-like neutralino.

If the stau decays outside the detector, the most restrictive bounds on the parameter space come from CHAMPs. Though bounds on stable staus
themselves are not
that restrictive, the overall production cross section limit for staus is dominated by production from gluino and chargino cascades.  Tevatron places
a bound on weakly produced CHAMPS at 10 $fb$.  Production from gluinos can be safely supressed for gluinos above 500 GeV.  The most restrictive bound
however comes from chargino pair production, which requires chargino masses of 250 GeV for sufficient suppression.  This beats all other bounds and
forces
us into a corner of parameter space  similar to that of the case for one messenger with non-prompt NLSP decay.

In the case of prompt decay, GMSB searches place a lower bound on right handed staus of 82.5 GeV in case of stau NLSP.  In areas with bino-like NLSP
the 102.7 GeV chargino bound is the most restrictive.  In this case squark masses may be pushed down to about 450 GeV again with a spectrum qualitatively
like the case for minimal gauge mediation with multiple messengers.

Below in Table 3 is a typical spectrum for non-prompt decay of the NLSP
\begin{table}[h]
\begin{center}
\begin{tabular}{c|c}
sparticle & mass (GeV)  \\
\hline
$\tilde{\tau}$  & 124  \\
$\tilde{\nu}$  & 220 \\
$\tilde{q_{ul}}$  & 482 \\
$\mu$  & 327   \\
$\chi^{0}$  & 266  \\
$\tilde {g}$ & 685 \\
$ mz^2 / \delta_{mhu}^2$ & 20 \\

\end{tabular}
\caption{Spectrum for a 2-D subspace of 4 parameter GGM,  where $\Lambda_3 = $ 40 TeV $\Lambda_2 =$ 58 TeV with  Higgs soft masses set to the GGM prediction.}
\label{tab:gauginobounds}
\end{center}
\end{table}

\section{Three Parameters}

\begin{figure}[h]
\centerline{\includegraphics[width=10 cm]{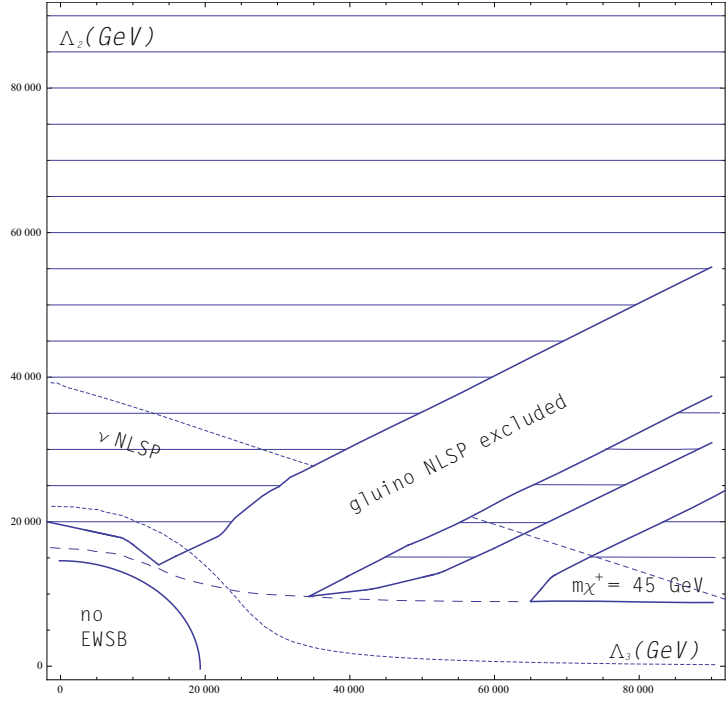}}
\caption{Plot of $\Lambda_3$ vs $\Lambda_2$ with various constraints.  $\Lambda_E$ = 80 TeV and the SUSY breaking scale is high}  
\label{fig:3lowpgm}
\end{figure}

\begin{figure}[h]
\centerline{\includegraphics[width=10 cm]{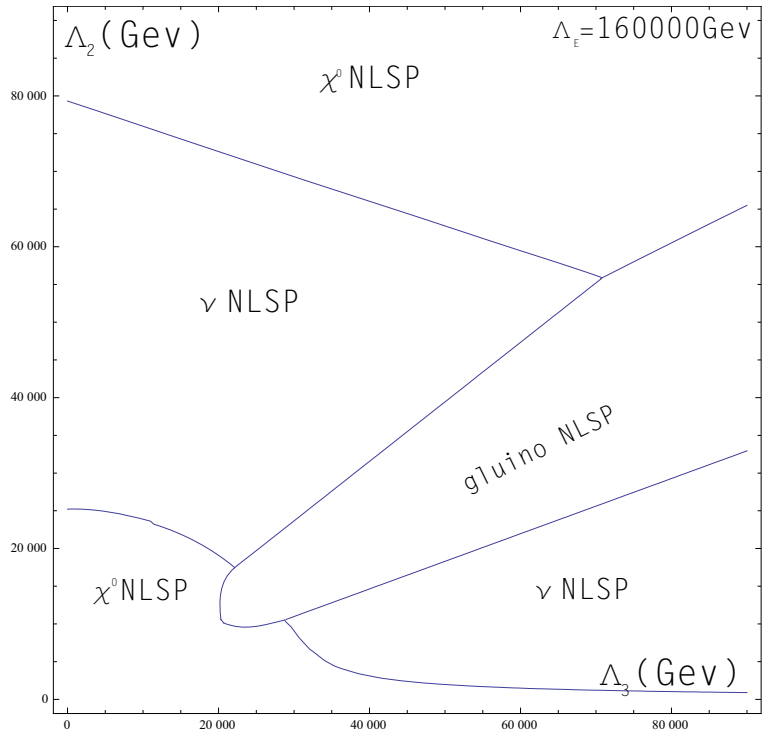}}
\caption{Plot of $\Lambda_3$ vs $\Lambda_2$ for regions of differing NLSP.  $\Lambda_E$ = 160 TeV and the SUSY breaking scale is high}
\label{fig:3highpgm}
\end{figure}

Three parameter GM may be achieved with a set of messengers in a $10$, $\overline{10}$ representation of SU(5).  The relations for three parameter GM are
 given by
\be
m_{\tilde{g}}=\frac{\alpha_3}{4\pi}(2\Lambda_2+\Lambda_3);m_{\tilde{w}}=\frac{\alpha_2}{4\pi}3\Lambda_2;m_{\tilde{b}}=\frac{\alpha_1}{4\pi}(\frac{8}{3}\Lambda_2+\frac{4}{3}\Lambda_3+2\Lambda_1)
\ee

\be
m_{\tilde{s}}^2=2(C_3 (\frac{\alpha_3}{4\pi})^2 \Lambda_Q^2 +C_2 (\frac{\alpha_2}{4\pi})^2 \Lambda_w^2 +\frac{Y}{2}^2(\frac{\alpha_1}{4\pi})^2 \Lambda_Y^2
\ee
where $\Lambda_Q^2 = (2\Lambda_2^2+\Lambda_3^2)$, $\Lambda_w^2 = 3\Lambda_2^2$ and $\Lambda_Y^2= (\frac{8}{3}\Lambda_2^2+\frac{4}{3}\Lambda_3^2 + 2\Lambda_1^3)$

Here, with just three parameters, spectra diverge from the qualitative features of minimal gauge mediation.  The gaugino masses are now independent.
It is possible to make the gluino the NLSP by canceling the scales $\Lambda_3$ and $\Lambda_2$ while the bino may be made heavy by
choosing large $\Lambda_E$.   The choice of large $\Lambda_E$ also compresses the mass difference between sleptons and squarks. For regions where
$\Lambda_2$ is sufficiently low compared to the other scales, the sneutrino may be the NLSP. The analysis that follows scans the two dimensional
subspace of 3 parameter GM fixing the scale $\Lambda_E$ at
the relatively large value of 80 TeV.

For the case of high SUSY breaking scale, the gluino can be made to be the NLSP.  However, it is too light ($<$ 150 GeV) to be allowed by
stopped gluino constraints.  Elsewhere in parameter space, the bino remains the NLSP, except where $\Lambda_2$ is sufficiently smaller than the other scales
 such that the sneutrino is the NLSP.  This occurs for low values of $\Lambda_2$ except near the electroweak trough.
 For these large values of $\Lambda_E$ the U(1) mass contributions to scalars are relatively heavy.  In low $\Lambda_2$ regions the charged
 fermions are heavier than the charginos.  However, the sneutrinos are light in regions of low $\Lambda_2$ therefore
 the chargino mass bounds are not strong.  Again there is a small window of wino-like light neutralino.  However,
this window is ruled out as the NLSP in the area is not the neutralino but the sneutrino.  Finally, some part of parameter space
is ruled out by model independent gluino to jets plus missing energy searches.  Figure 2 shows
a scan of the high SUSY breaking scale three parameter space.


Constraints on promptly decaying sparticles in this 3 parameter case are similar.  With the same choice of $\Lambda_E$, gluino NLSPs are still ruled out,
this time by monojet analyses.  The chargino mass bound is increased to 229 GeV for bino-like NLSP.  In the case of a sneutrino NLSP, the chargino decays
in the lepton plus missing energy channel and the chargino mass constraint is much weaker.  Unlike the minimal and two parameter case, prompt decay in the
three parameter case is almost as unconstrained as non-prompt decay.

In both cases, since U(1) contributions to scalar masses can be made very large.  The possibility exists for an extremely compressed spectrum,
about a factor of 3 compression from the minimal scenario.  However the compression results from raising charged slepton masses rather than from
decreasing squark masses, therefore tuning remains at about a part in 20.  Below, Table 4 shows a typical point in parameter space for 3 parameter gauge mediation in the case of non-prompt decay.

\begin{table}[h]
\begin{center}
\begin{tabular}{c|c}
sparticle & mass (GeV)  \\
\hline
$\tilde{\tau}$  & 223  \\
$\tilde{\nu}$  & 184 \\
$\tilde{q_{ul}}$  & 487  \\
$\mu$  & 372  \\
$\chi^{+}$  & 207  \\
$\chi^{0}_1$  & 201  \\
$\tilde {g}$ & 413 \\
$ mz^2 / \delta_{mhu}^2$ & 23 \\

\end{tabular}
\caption{Typical spectrum, this point is for $\Lambda_3$ 10 TeV $\Lambda_2$ 28 TeV with no extra contribution to Higgs soft masses.}
\label{tab:gauginobounds}
\end{center}
\end{table}

If the scale $\Lambda_E$ is made even larger in comparison to the other scales the phenomenology becomes even more complex.  Figure 4 shows a survey
of possible NLSPs for the high SUSY breaking scale benchmark and the choice $\Lambda_E = $ 160 TeV.  In this case the region over which the gluino is
the NLSP is increased.  This time there is a small strip of parameter space where the gluino is still the NLSP and is more massive than the bound demanded by
the stopped gluino searches.  Here the gluino may be up to around 360 GeV.  In addition since $\Lambda_E$ is larger than $\Lambda_2$ over much of parameter
space, the region in which the sneutrino is the NLSP increases.  In the region where the gluino is the NLSP, the squarks are quite heavy. This is
because while mass parameters cancel for the gluino, they add in quadrature for the squarks, and squark masses end up around 1.5 TeV.
This region is highly tuned and the spectrum is quite hierarchical, however it offers the possibility of a SUSY spectrum where everything decays to
multi-jets plus missing energy.   In regions with a sneutrino NLSP, the tuning in the spectrum can remain quite reasonable.  The spectrum may become
exceedingly compressed with right handed squarks and sleptons around the same mass, while certain decays - for example that of the sleptons - become very nonstandard.

\section{Discussion}
We have seen that GGM can produce a variety of non-standard results.  Scanning over just 3 GGM parameters qualitative features of the spectrum may change drastically;
 NLSPs may vary from wino to bino to sneutrino to stau to gluino.  Several features seem to be generic over the space.
 First of all getting a gluino NLSP is difficult.  For low numbers of parameters prompt decay of the NLSP is
 highly constrained, while parameter space is quite open as we reach the 3 parameter case.  In addition, though spectra may be compressed, squarks
 usually remain the heaviest sparticles.  An exception is the three parameter space with very large U(1) mass parameter, where squarks and sleptons may
 have similar masses.  In general, tuning in the multi-parameter space may be reduced to one part in 20, less than half the tuning of the minimal case; in some instances
 tuning may be as low as a part in 10.

 Many possibilities exist for interesting analyses that could further restrict this space.  Examples include: a full trilepton analysis for the case of
  non-prompt decay
 in minimal and multiparameter space; a model independent squark search that did not rely on gaugino mass unification;
 a gluino-bino exclusion analysis for the case of light squarks; analyses that explore sparticle bounds in cases of low mass splitting
 between the chargino and a wino-like NLSP; and slepton and chargino mass bounds in the case of a sneutrino NLSP.

{\bf Acknowledgments}

This work was supported in part by DOE grant number DE-FG03-92ER40689. I would like to thank Michael Dine, Howie Haber, and Andrew Blechman for helpful
discussions.

\end{document}